\documentclass[conference]{IEEEtran}
\usepackage{}
\usepackage{amssymb}
\usepackage{amsmath}
\usepackage{amsfonts}
\usepackage{graphicx, subfigure, cite, verbatim, url, amssymb, amsmath}
\usepackage{pifont}

\input ./psfig.sty

\def\captionstyle{}
\def\boxcaptionstyle{\raggedright}

\def\maxfigfraction{.6}

\newdimen\figboxmargin
\newdimen\figboxhang
\def\DVIscaling{1}
\def\globalscaling{1}
\def\figuredirectory{./figures}
\let\boxer=\llboxer

\def\missingfigure#1{\hbox{Missing figure #1.ps}}

\catcode `\@=11

\newbox\figurebox

\def\figbox{
\@ifnextchar[{\figboxaux}{\figboxaux[htb]}}

\long\def\figboxaux[#1#2]#3#4#5#6{
\writepict{{#3}{#4}{#5}{#6}}
\setbox\figurebox\hbox{#3}%
\if l#1\tryleftbox{#4}{#5}{#6}%
\else
\if r#1\tryrightbox{#4}{#5}{#6}%
\else
\if *#1\checktwocoloptions#2]{\box\figurebox}{#4*}{#5}{#6}%
\else\tryonecol[#1#2]{#4}{#5}{#6}%
\fi
\fi
\fi\ignorespaces}

\long\def\tryleftbox#1#2#3{
\ifdim\wd\figurebox>\maxfigfraction\columnwidth \tryonecol[htb]{#1}{#2}{#3}%
\else\leftbox{\captionbox{\box\figurebox}{#1}{#2}{#3}}\fi}

\long\def\tryrightbox#1#2#3{
\ifdim\wd\figurebox>\maxfigfraction\columnwidth \tryonecol[htb]{#1}{#2}{#3}%
\else\rightbox{\captionbox{\box\figurebox}{#1}{#2}{#3}}\fi}

\def\checktwocoloptions{
\@ifnextchar]{\floatbox[htb}{\floatbox[}}

\long\def\tryonecol[#1]#2#3#4{
\ifdim\wd\figurebox>\columnwidth \floatbox[#1]{\box\figurebox}{#2*}{#3}{#4}%
\else\floatbox[#1]{\box\figurebox}{#2}{#3}{#4}\fi}

\long\def\floatbox[#1]#2#3#4#5{%
\begin{#3}[#1]
\hbox to \hsize{\hfil#2\hfil}
\captionandlabel{#3}{#4}{#5}
\end{#3}
}

\long\def\captionbox#1#2#3#4{
\setbox\figurebox\hbox{#1}%
\parbox[t]{\wd\figurebox}{%
\bigskip\box\figurebox
\let\captionstyle=\boxcaptionstyle
\captionandlabel{#2}{#3}{#4}
\bigskip
}}

\def\captionandlabel#1#2#3{
\def\testit{#3}%
\ifx\testit\empty\else
\writecapt{{#1}{#2}{#3}}
\captypeunstarred#1*.
\getcaption#3\endc@ption
\def\testit{#2}
\ifx\testit\empty\else\label{#2}\fi
\fi}

\def\captypeunstarred#1*#2.{
\def\@captype{#1}}

\def\getcaption{\@ifnextchar[{\getcaptwo}{\getcapone}}
\long\def\getcapone#1\endc@ption{\caption[#1]%
{\def\baselinestretch{1}\Large\normalsize\captionstyle\ignorespaces #1}}
\long\def\getcaptwo[#1]#2\endc@ption{\caption[#1]%
{\def\baselinestretch{1}\Large\normalsize\captionstyle\ignorespaces #2}}

\newdimen\figboxht
\newcount\figboxn

\newcount\figboxlines
\newdimen\figboxwid

\newif\ifisleftbox

\long\def\leftbox#1{%
\setbox\figurebox\hbox{#1}\global\isleftboxtrue
\startmarginbox
\vadjust{\smash{\rlap{\hskip\hsize\hskip\figboxhang
\llap{\raise.7\baselineskip\box\figurebox\hskip\rightskip}}}}%
\endmarginbox%
}

\long\def\rightbox#1{%
\setbox\figurebox\hbox{#1}\global\isleftboxfalse
\startmarginbox
\smash{\llap{\raise.7\baselineskip\box\figurebox\hskip\figboxmargin}}%
\endmarginbox%
}

\def\startmarginbox{%
\ifvmode\passpict\let\endmarginbox=\indent
\else\message{WARNING: marginbox in not in vmode}\hfilneg\ \passpict
\let\endmarginbox=\relax\fi
\figboxht=\dp\figurebox
\advance\figboxht by 1.3\baselineskip
\vskip\figboxht\penalty-300\vskip-\figboxht
\divide\figboxht by\baselineskip
\global\figboxlines=\figboxht
\global\figboxwid=\wd\figurebox
\global\advance\figboxwid by \figboxmargin
\global\advance\figboxwid by -\figboxhang
\setmypar\noindent}
\def\addlines#1{\global\advance\figboxlines by #1\myparshape}
\def\zerolines{\origpar\global\figboxlines=0\myparshape}

\def\passpict{\par\ifnum\figboxlines>1\vskip\figboxlines\baselineskip
\zerolines\fi}

\def\emptybox#1#2{\hbox to #1{\vbox to #2{\vss}\hss}}

\global\let\origpar=\@@par
\global\let\dopar=\origpar
\global\def\@@par{\dopar}
\@setpar{\dopar}

\def\setmypar{\global\let\dopar=\mypar
\global\prevgraf=0\myparshape}

\def\mypar{\origpar\global\advance\figboxlines by -\prevgraf%
\global\prevgraf=0\myparshape}

\def\myparshape{\relax%
\ifnum\figboxlines>1\theparshape \else
\global\hangindent=0pt\global\hangafter=1
\global\let\dopar=\origpar\fi}

\def\theparshape{%
\ifisleftbox\global\hangindent=-\figboxwid 
\else\global\hangindent=\figboxwid \fi
\global\hangafter=-\figboxlines \global\advance\hangafter by 1%
}

\def\definefnum#1{
\def\fnum@figure{Figure \ref{#1}}%
\def\fnum@table{Table \ref{#1}}%
\def\fnum@code{Algorithm \ref{#1}}%
}

\def\writepict#1{}
\def\writecapt#1{}

\def\journalpicts#1{
\newwrite\pictfile
\newwrite\captfile
\openout\pictfile\jobname.pic
\openout\captfile\jobname.cap
\gdef\writepict##1{\unexpandedwrite\pictfile{\doit##1}}%
\gdef\writecapt##1{\unexpandedwrite\captfile{\doit##1}}%
\global\let\ENDdocument=\enddocument
\gdef\enddocument{\DOjournalpicts{#1}\ENDdocument}
}

\def\DOjournalpicts#1{{%
\def\writepict##1{}\closeout\pictfile
\def\writecapt##1{}\closeout\captfile
\@fileswfalse
\onecolumn
\def\globalscaling{#1}
\def\doit##1##2##3##4{
\figboxaux[t]{\hss##1\hss}{##2}{}{}%
\vspace*{1in}
\definefnum{##3}
\captionandlabel{##2}{}{##4}
\clearpage}%
\input\jobname.pic
\def\doit##1##2##3{
\definefnum{##2}
\captionandlabel{##1}{}{##3}}%
\raggedright\let\captionstyle=\raggedright
\def\@makecaption##1##2{##1: ##2\par}
\input\jobname.cap
}}

\def\llboxer#1{\vbox to \figboxht{\vfil\hbox to \figboxwid{#1\hfill}}}
\def\lcboxer#1{\vbox to \figboxht{\vfil\hbox to \figboxwid{\hfill#1\hfill}}}
\def\oldboxer#1{\vbox to \figboxht{\vfil
		      \hbox to \figboxwid{\hfill\llap{#1\hskip4.25in}\hfill}}}
\def\ulboxer#1{\vbox to \figboxht{\hbox to \figboxwid{#1\hfill}\vfil}}
\def\ccboxer#1{\vbox to \figboxht{\vfil
			\hbox to \figboxwid{\hfill#1\hfill}\vfil}}

{\catcode`\p=12\catcode`\t=12
\gdef\removedimen#1pt{#1}}

\def\defscaled#1#2{#2=\DVIscaling#2%
\xdef#1{\expandafter\removedimen\the#2}}

\def\DVIspace{ }
\newdimen\hscalefactor
\newdimen\vscalefactor

\def\scale#1{\horizscale{#1}\vertscale{#1}}
\def\horizscale#1{\hscalefactor=#1\hscalefactor\figboxht=#1\figboxht}
\def\vertscale#1{\vscalefactor=#1\vscalefactor\figboxwid=#1\figboxwid}

\def\boxps{%
\@ifnextchar[{\boxpsaux}{\boxpsaux[\relax]}}

\def\boxpsaux[#1]#2#3#4#5{%
{\figboxwid#4\figboxht#5\hscalefactor=1pt\vscalefactor=1pt%
\scale{#3}%
\scale{\globalscaling}%
#1%
\defscaled\DVIhscale\hscalefactor
\defscaled\DVIvscale\vscalefactor
\boxer{\includegraphics{\figuredirectory/#2.ps}}}%
}

\newread\Epsffilein
\newif\ifEpsffileok
\newif\ifEpsfbbfound

\newdimen\pspoints
\divide\pspoints by 72

\def\boxeps{%
\@ifnextchar[{\boxepsaux}{\boxepsaux[\relax]}}
\def\boxepsaux[#1]#2#3{%
\openin\Epsffilein=\figuredirectory/#2.ps 
\ifeof\Epsffilein\message{I couldn't open \figuredirectory/#2.ps }%
\missingfigure{#2}
\else
   {\Epsffileoktrue\Epsfbbfoundfalse
    \catcode`\%=11 \catcode`\\=11
    \catcode`\{=11 \catcode`\}=11
    \catcode`\$=11 \catcode`\^=11
    \catcode`\&=11 \catcode`\#=11
    \catcode`\~=11 \catcode`\_=11
    \loop
       \read\Epsffilein to \Epsffileline
       \ifeof\Epsffilein\Epsffileokfalse\else
          \expandafter\Epsfaux\Epsffileline . .\\%
       \fi
   \ifEpsffileok\repeat
   \ifEpsfbbfound
	\figboxht=\Epsfury\pspoints
	\advance\figboxht by-\Epsflly\pspoints
	\figboxwid=\Epsfurx\pspoints
	\advance\figboxwid by-\Epsfllx\pspoints
   \else
	\message{No bounding box comment in \figuredirectory/#2.ps }%
	\figboxwid=2in\figboxht=1in%
   \fi%
   \immediate\closein\Epsffilein
   \hscalefactor=1pt\vscalefactor=1pt%
   \scale{#3}%
   \scale{\globalscaling}%
   #1%
   \defscaled\DVIhscale\hscalefactor
   \defscaled\DVIvscale\vscalefactor
   \hscalefactor=-\Epsfllx\hscalefactor
   \hscalefactor=1.00375\hscalefactor
   \defscaled\DVIhoffset\hscalefactor
   \vscalefactor=-\Epsflly\vscalefactor
   \vscalefactor=1.00375\vscalefactor
   \defscaled\DVIvoffset\vscalefactor
   \llboxer{\includegraphics{\figuredirectory/#2.ps}} }%
\fi
}%
{\catcode`\%=11 \global\let\Epsfpar=\par
\global\let\Epsfpercent=
\long\def\Epsfaux#1#2 #3\\{\relax\ifx#1\Epsfpercent
   \def\testit{#2}\ifx\testit\Epsfbblit
      \Epsfsize #3 . . . .\\%
      \global\Epsffileokfalse
      \global\Epsfbbfoundtrue
   \fi\else\ifx#1\Epsfpar\else\global\Epsffileokfalse\fi\fi}%
\def\Epsfsize#1 #2 #3 #4 #5\\{\global\def\Epsfllx{#1}\global\def\Epsflly{#2}%
   \global\def\Epsfurx{#3}\global\def\Epsfury{#4}}%

\catcode`\@=12 			

\def\pic#1;#2;#3;#4\par{\picsc#1;#2;#3;1;#4\par}

\def\picsc#1;#2;#3;#4;#5\par{
\figbox[htb]{\boxeps{#1}{#4}
}{figure}{#1}{%
#5}}

\def\mpic#1;#2;#3;#4\par{\mpicsc#1;#2;#3;1;#4\par}

\def\mpicsc#1;#2;#3;#4;#5\par{
\figbox[l]{\boxeps{#1}{#4}
}{figure}{#1}{%
#5}}

\def\BibTeX{{\rmfamily B\kern-.05em{\scshape i\kern-.025em b}\kern-.08em \TeX}}

\newcommand{\ls}[1]  
   {\dimen0=\fontdimen6\the=#1\dimen0
    \advance\lineskip.5\fontdimen5\the\lineskip-\dimen0
    \lineskiplimit=.9\lineskip
    \baselineskip=\lineskip
    \advance\baselineskip\dimen0
    \normallineskip\lineskip
    \normallineskiplimit\lineskiplimit
    \normalbaselineskip\baselineskip
    \ignorespaces
   }

\begin{document}
%
\title{Fountain Uncorrectable Sets and Finite-Length Analysis}

\author{
\IEEEauthorblockN{Wen Ji$^{1}$, Bo-Wei Chen$^{2}$, and Yiqiang Chen$^{1}$}
\IEEEauthorblockA{$^{1}$Beijing Key Laboratory of Mobile Computing and Pervasive Device\\ Institute of Computing Technology, Chinese Academy of Sciences, Beijing 100190, P.R. China\\
$^{2}$Electrical Engineering Department, Princeton University, NJ, 08544, USA\\
E-mail: jiwen@ict.ac.cn; dennisbwc@gmail.com; yqchen@ict.ac.cn.
}}

\IEEEcompsoctitleabstractindextext{%
\begin{abstract}
Decoding performance of Fountain codes for the binary erasure channel (BEC) depends on two aspects. One is the essential code structure, on which stopping set analysis operates. The other is the effect from the channel characteristic, which is difficult to give a precise estimation. To tackle these problems, in this paper, we propose a solution to analyzing the performance of Fountain codes based on the uncorrectable set. We give the condition for Fountain decoding failure over the BEC. Then, we conduct the analysis of uncorrectable set on Fountain codes. Finally, we combine the stopping set and the uncorrectable set to provide the integrated analysis on the performance of Fountain codes for BEC.
\end{abstract}

}

\maketitle

\IEEEdisplaynotcompsoctitleabstractindextext

\IEEEpeerreviewmaketitle

\section{Introduction}

\IEEEPARstart For binary linear codes, the decoding performance of
belief propagation (BP)-based iterative decoding is dominated by
stopping sets over the binary erasure channel (BEC). Stopping sets were firstly introduced for the
analysis of low-density parity-check (LDPC) codes over BECs \cite{Di-Proietti-TIT-02-06}.
It was shown that the iterative decoder failed to decode to a
codeword if and only if the set of erasure positions was a superset
of some stopping set in the Tanner graph during decoding. In
particular, the number and the size of stopping sets is important
for determining the performance of iterative decoders. Stopping
sets in a small size for the BEC can lead
to small Hamming distance.
The success of stopping sets in analyzing LDPC codes has created a paradigm for researchers to analyze the other codes. For example, Rosnes and Ytrehus introduced the concept of stopping sets to
analyze turbo decoding and proposed turbo stopping set \cite{Rosnes-Ytrehus-TIT-07-11}.
Abdel-Ghaffar and Weber derived an equation based on the number of
stopping sets for a full-rank parity-check matrix of the Hamming
code\cite{Abdel-Weber-TIT-07-09}. Tuvi examined the stopping redundancy
Reed-Muller codes \cite{Tuvi-TIT-06-11}. Wadayama presented the stopping set
of redundant random ensembles \cite{Wadayama-TIT-08-11}.

Recently,
much attention has been given to a class
of error-control codes, Fountain codes, due to their excellent performance, especially in erasure channels and the simplicity of encoding and decoding.

Three typical examples of rateless codes were developed based upon the Fountain codes: Luby Transform (LT) codes\cite{Luby-FOCS-02}, Raptor codes\cite{Shokrollahi-TIT-06-06}, and Online codes. As LT codes own the basic structure of Fountain code family, many studies on error analysis were conducted based on LT codes.

For instance, the error analysis reported in \cite{Karp-Luby-CISIT-04} gave a basic result depending on the exact calculation of the error probability. The works in \cite{Abdulhussein-Oka-TCL-08-06} and \cite{Orozco-Yousefi-QBSC-10} respectively developed stopping criterions so as to detect the earlier decoding termination with a lower cost.

Although the error-control mechanism in Fountain codes facilitated error analysis, two major factors in Fountain codes on the BEC still affect decoding performance. One is the essential codes structure, on which stopping set analysis operates. The other is the effect from the channel characteristic, which has not been effectively resolved yet. Current finite length analysis nonetheless still focused on the former problem -- the error-prone structures of codes. It is much more difficult to give a precise estimation of error-prone patterns.

As Fountain code family belongs to nonsystematic codes, which are different from the existing families like LDPC and Turbo, the conventional stopping set is not applicable. To overcome such a problem, in this study, we focus on the performance analysis when output nodes are erased. We introduce uncorrectable set in Fountain codes in order
to analyze the decoding performance of Fountain codes over the BEC. Furthermore, we also provide the concept of uncorrectable set and analyze
the probability of bit erasure of Fountain codes over the BEC in average.%

The rest of this paper is organized as follows. Section II briefs the LT code. Section III then describes the Foutain uncorrectable set. Next, Section IV shows the probability of bit erasure followed by the integrated performance analyze in Section V. Conclusions are finally drawn in Section VI.

\section{Preliminaries}\label{sec_2}

\subsection{Principle of LT codes }
Fountain codes include three typical classes: Luby Transform (LT)
codes, Raptor codes, and Online codes. Among these, LT codes is the
basic to construct other families. LT code retains good performance
of random linear fountain code, while drastically reduces the
complexities both in encoding and decoding process. During encoding, LT
divides the uncoded message into $k$ blocks with roughly equal
length. The degree $d$ $(1\leq d\leq k)$ of the next packet is
is randomly chosen. Accordingly, $d$ input symbols are chosen
uniformly at random.

Let $G$ denote a generation matrix for a length given LT code. The
encoding can be represented by:

\begin{equation}
t_i=\sum_{j=1}^{k}x_j\cdot G_{ji}
\end{equation}where $n$ is the code length, $k$ is the length of the input symbol, $t_i$
denotes the $i$th of encoded symbol, $x_j$ denotes the $j$th of encoding symbol. Without loss of generality, this paper considers the symbol is binary.

\subsection{The graph representation of LT codes}
The parity-check matrix $H$ can also be represented by a bipartite
graph $\mathcal {G}=(\mathcal {V}\cup \mathcal {C},\mathcal {E})$, where the set of variable nodes $\mathcal {V}$
represents the codeword symbol and the set of check nodes $C$
represents the set of parity-check constraints satisfied by the
codeword bits, and edges $\mathcal {E}\subset \{(v,c)|v\in \mathcal {V}, c\in \mathcal {C}\}$. First,
let us briefly review conventional stopping sets in LDPC codes. The
concept of stopping sets is proposed based on Tanner graph. A
stopping set $S$ in a code is a subset of the variable nodes in a
Tanner graph for $\mathbb{C}$ such that all the neighbors of $S$ are
connected to $S$ at least twice.

For a given matrix $G_{k, n}$, let  $X=(x_1,x_2,...,x_k)$ denote the
encoding symbols. Let $T = (t_1,t_2,...,t_n)$ denote the codeword.
Then, $X\cdot G_{k, n}=T$. In general case, the relation $G_{k,n}H^{\textrm{T}}=0$ is adopted to computer the parity-check matrix $H$.

For binary linear systematic code, parity-check matrix $H$ of LDPC
is obtained according to $G_{k,n}H^T=0$. The matrix $H$ can verify
the estimation value of $X=(x_1,x_2,...,x_k)$ because LDPC is
systematic code; $T$ can be represented by
$T=(x_1,x_2,...,x_k,p_1,p_2,...,p_{n-k})$, where
$p_1,p_2,...,p_{n-k}$ denotes the parity bits. Thus, the encoding
bits $X$ are included in the transmitted bits $T$ and are sent to
the receiver.

However, LT codes are nonsystematic codes which only transmit parity
symbols. $T$ can be represented by $T=(p_1,p_2,...,p_{n-k})$. The
transmitted symbols do not include the encoding symbols $X$. Then, the
matrix $H$ deduced from $G_{k,n}H^T=0$ only verifies the transmitted
symbols $T$ but not to verify the encoding bits $X$. For the sake of
clarity, here we only concern the validity of encoding symbols
$X=(x_1,x_2,...,x_k)$ without caring for the transmitted symbols
$T=(t_1,t_2,...,t_n)$. Therefore, the conventional solution on
parity-check matrix $H$ must be changed in order to suitable to LT
codes.

We propose a method which can create the parity-check matrix of LT
codes: Since the transmitted bits are either lost or correct when
the code transmits on BEC, the all received bits are correct. Let
$P=(p_1,p_2,...,p_r)$ represent the received bits. The partitions of
matrix $G_{k,n}$ corresponding to $P=(p_1,p_2,...,p_r)$ make up of
the matrix $G_{k,r}$. There is,

\begin{equation}
X\cdot G_{k, r}=P.
\end{equation}

Let $\mathcal {G}\{k,\lambda (v),\rho (d)\}$ denote a Fountain code
ensemble, where $k$ is input symbol length, $\lambda (v)$ is the
degree distribution of input node, and $\rho (d)$ is the degree
distribution of output node. From the above analysis, the matrix
$G_{k,r}$ plays the role in the parity-check matrix which can verify the
encoding bits $X=(x_1,x_2,...,x_k)$ in Fountain codes. Hence, for a
particular code $G\in \mathcal {G}$ can also be represented by a
bipartite graph $\mathcal {G}=\{\mathcal{V}\cup \mathcal{C},\mathcal{E}\}$, where the set of variable nodes
$\mathcal{V}$ represents $k$ input nodes, corresponding to the input symbols.
The set of check nodes $\mathcal{C}$ represents the set of parity-check
constraints satisfied by the input symbols, corresponding to the
output symbols,and edges $\mathcal{E}\subset \{(v,c)|v\in \mathcal{V}, c\in \mathcal{C}\}$.

\section{Fountain Uncorrectable Set}

In this section, we analyze the decoding performance of Fountain
code over BEC. It is known that the length of output symbols directly reflects
the performance of iterative decoding algorithms. According to the
above analysis, we build the Tanner graph of $G_{k,n}$, as Shown in
Fig.1. Circular nodes correspond to the input symbols, and the
rectangular nodes correspond to the output symbols. There exists an
edge between the input symbol and output symbol if and only if
$a_{ij}=1$, where $a_{ij}$ denotes the element of generator matrix
in the $i$th row and $j$th column.

We define Fountain uncorrectable set as follows.

\textbf{Definition 1.} An uncorrectable set $\mathcal {U}$ in Fountain codes represents
a subset $\mathcal {V}$ of information nodes. The nodes directly
connected to $\mathcal {V}$ will be erased.

As shown in Fig.1, the different line type expresses an uncorrectable set. For the code in Fig.1, if only $c_1$ is erased, the maximal uncorrectable set is $\mathcal {U}=\{{\O}\}$. If $c_1, c_2$, and $c_3$ are deleted, it means that the connected $v_1$ and $v_4$ cannot be decoded successfully. Accordingly, the uncorrectable set is $\mathcal {U}=\{v_1,v_4\}$.

\textbf{Properties.} An uncorrectable set has the following properties:

 1) The union of uncorrectable sets is also an uncorrectable set.

 2) Each erasure pattern contains a unquie maximal uncorrectable set which
 might be an empty set.

\section{Symbol Erasure Probability}

For a particular code $G$ in a given ensemble $\mathcal
{G}(k,\lambda (v),\rho (d))$, let $P_b(G,\varepsilon)$ denote the
expected bit erasure probability if $G$ is used to transmit over a
BEC with erasure probability $\varepsilon$. Let $E_{\mathcal
{G}(k,\lambda (v),\rho (d))}[P_b(G,\varepsilon)]$ denote the probability of corresponding ensemble average bit erasure. Assuming the
number of erasure bits is $|e|$, where $e$ denotes the pattern of
erasure. There are $E(e)$ output node sockets in some arbitrary but
fixed way with elements from the set $e$. Similarly, there are also
input node sockets in some arbitrary but fixed way with elements
from the set $V(e)$. The element of $V(e)$ cannot be recovered. As
shown in Fig.2, the rectangular nodes with black correspond to the
$|e|$ lost output symbols. Circular nodes with black correspond to
the $V$ input symbols cannot be recovered because output symbols
incident upon them are all lost. Circular nodes with gray correspond
to the input symbols may be recovered because output symbols
incident upon them are not all lost.

\begin{figure}
\noindent\begin{minipage}{\linewidth}
\begin{minipage}[t]{.48\linewidth}
\centering\includegraphics[width=1.5in]{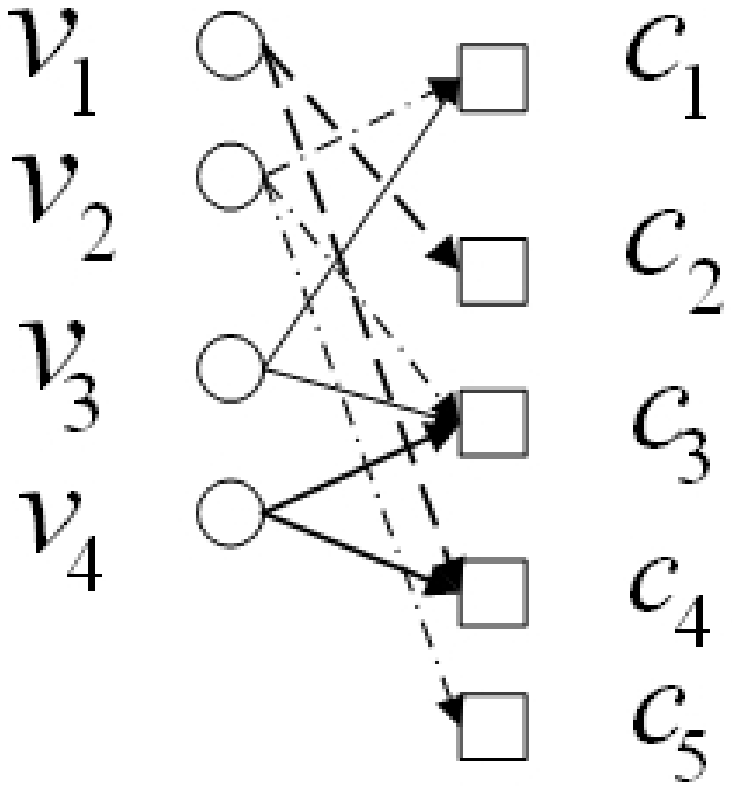}\caption{Fountain
uncorrectable set}\label{fig1}
\end{minipage}
\begin{minipage}[t]{.48\linewidth}
\centering\includegraphics[width=1.4in]{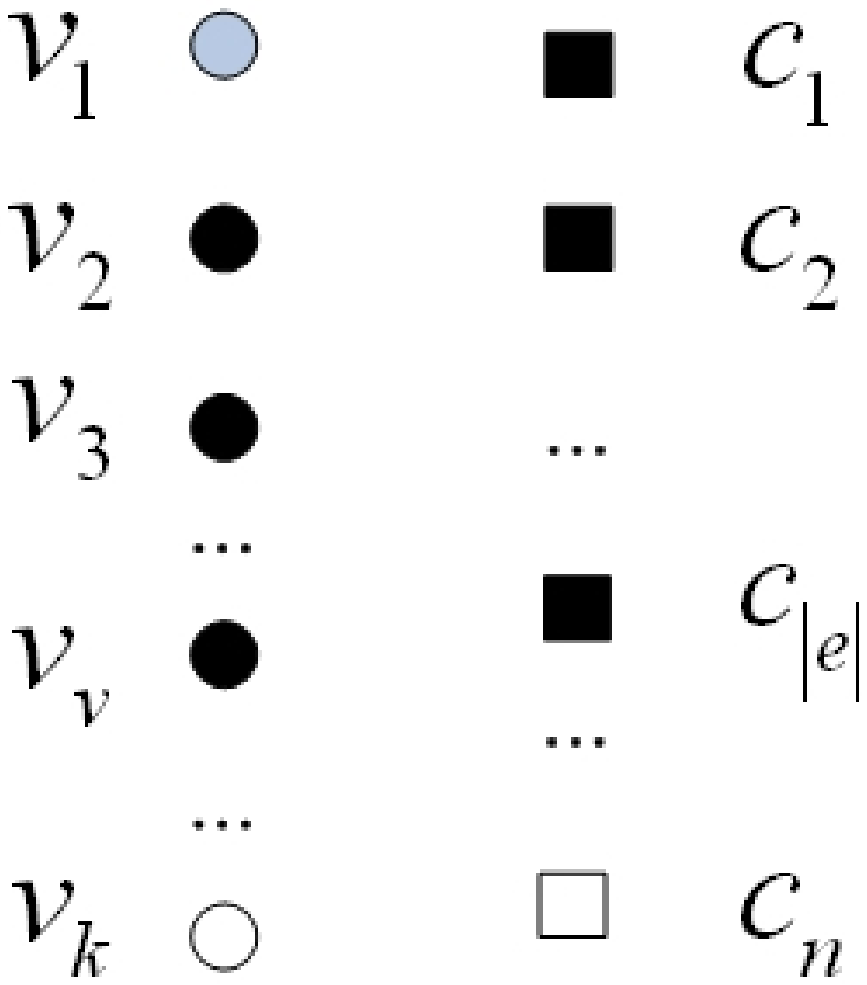} \caption{There are
$|e|$ output nodes lost, which lead to $V$ input nodes undecodable}
\end{minipage}
\end{minipage}
\vspace{-1em}
\end{figure}

When $|e|$ output nodes are lost, the edges incident upon them are
also lost. The following the number of edges connected to $|e|$
output nodes lost is computed.

\textbf{Theorem 1.} {\em The probability of the number of edges with
$L$ connected to the set $e$ in Fountain codes $\mathcal
{G}\{k,\lambda (v),\rho (d)\}$ is:
\begin{equation}\label{Theorem1}
\mathrm{coef}\bigg(\prod^{d_{\max}}_{i=1}(1+xz^i)^{\rho_in},x^{|e|} z^L
\bigg)\bigg/{n \choose |e|}
\end{equation}
}where $\mathrm{coef}(f(x),x^i)$ denotes the coefficient of $x^i$ in the
polynomial $f(x)$, and $d_{\max}$ denotes the maximal degree of output
nodes. Since
$\mathrm{coef}\big(\prod^{d_{\max}}_{i=1}(1+xz^i)^{\rho_in},x^{|e|} z^L
\big)$ is the number of sets with $|e|$ output nodes and $L$ edges
incident upon them, the total numbers
for selecting the pattern of erasure set $e$ are $\binom{n}{|e|}$. Combing the above
equation, then the edge distribution connected to $e$ is
(\ref{Theorem1}).

Now, we consider the number of input nodes incident upon $E=|E(e)|$.

\textbf{Theorem 2. } {\em The average bit erasure probability for
Fountain ensembles $\mathcal {G}\{k,\lambda (v),\rho (d)\}$ when
transmitting over a BEC with erasure probability $\varepsilon$ is
\begin{equation}\label{Theorem2}
\begin{split}
&E_{\mathcal
{G}(k,\lambda (v),\rho (d))}[P_b(G,\varepsilon)]\\
&=\sum_{|e|}\dbinom{n}{|e|}\varepsilon^{|e|}(1-\varepsilon)^{n-|e|}\\
&\times\sum_{L=1}^{E}\mathrm{coef}\big(\prod^{d_{\max}}_{i=1}(1+xz^i)^{\rho_i
n},x^{|e|} z^L\big)\bigg/{n \choose |e|}\\
&\times\sum^k_{V=1}\frac{V}{k}\times P(e,L,V)\\
 \end{split}
\end{equation}
}where $P(e,L,V)$ denotes the probability of the uncorrectable set when the maximum size of uncorrectable edges reaches $L$, and the maximal size of uncorrectable set is equal to $V$.

\begin{proof} Note that for Fountain ensembles $\mathcal {G}\{k,\lambda
(v),\rho (d)\}$, if all edges incident upon an input node belong
to the edges connected to $e$, the uncorrectable set of this input node
is lost. Hence, this input node cannot be recovered.

Assume that the set of $V$ nodes connected with $L$ edges forms the maximal uncorrectable set. Hence, the number of sets with $V$ input nodes and $L$ $(0\leq L\leq
E)$ edges incident upon them is
\begin{equation}
M_1(k,L,V)=\sum_{l\leq L}\mathrm{coef}\bigg(\prod^{v_{\max}}_{j=1}(1+yz^j)^{\lambda _j
k},y^V z^L \bigg)(l)!.
\end{equation}

Let $T(k,n)$ denote the number of the all maps with $k$ input nodes
connected to $n$ output nodes. It is
\begin{equation}
T(k,n)=(\sum^{v_{\max}}_{j=1}j\times \lambda _j \times k)!.
\end{equation}

Let $\mathcal {U}$ be an uncorrectable set if it contains a nonempty subset of the variable nodes such that any regular check node $\verb"c"$, which is connected to $\mathcal {U}$, is connected to $\mathcal {U}$ at least twice. Obviously, there is $\mathcal {U}\subseteq \mathcal {V}$, where $\mathcal {V}$ is the set of variable set.
Let $\mathcal {L}$ be the set that any check node, which is connected to $\mathcal {L}$ but not to $\mathcal {U}$, is connected to $\mathcal {L}$ at least twice. There is $\mathcal {L}\subseteq \mathcal {V}\setminus \mathcal {U}$. Let $\mathcal {K}$ be the maximal uncorrectable set. If every check node that is connected to $\mathcal {L}$ but not to $\mathcal {U}$ at least twice, there is $\mathcal {L}=\mathcal {K}\setminus \mathcal {U}$. If $\mathcal {V}\setminus \mathcal {U}$ does not contain a subset $ \mathcal {L}$ with the property that every check node with is connected to $\mathcal {L}$ at least twice.
Define the functions $Q(k,L,V)$, $N(k,L,V)$ and $M(k,L,V)$ by the recursions
\begin{equation}
Q(k,L,V):=\sum_{V>0} M(k,L,V)
\end{equation}
\begin{equation}
N(k,L,V):=T(k,n)-Q(k,L,V)
\end{equation}
\begin{flalign}
M(k,L,V):=M_1(k,L,V)\cdot N(k-V, E-L,0)
\end{flalign}where $M(k,L,V)$ is the number of maximal uncorrectable set $V$ with $E$ erasure edges. $N(k-V, E-L,0)$ denote the number which the remaining $k-V$ variable nodes with the remaining $E-L$ edges does not contain the uncorrectable set. And there are $k-V$ variable nodes in $\mathcal {V}\setminus \mathcal {U}$ and there are $E-L$ check nodes which are not neighbors of $\mathcal {U}$. We have
\begin{flalign}\nonumber
M(k,L,V)=&\Bigg(\sum_{l\leq L}\mathrm{coef}\bigg(\prod^{v_{\max}}_{j=1}(1+yz^j)^{\lambda _j
k},y^V z^L \bigg)(l)!\Bigg)\\
&\cdot N(k-V, E-L,0).
\end{flalign}

Then, the probability which the maximal uncorrectable set is equal to $V$ is $P(e,L,V)=\frac{M(k,L,V)}{T(k,n)}$.

It is easy to see that the probability is $\dbinom{n}{|e|}\varepsilon^{|e|}(1-\varepsilon)^{n-|e|}$ that pattern erasure is $e$.
The probability that $L$ edges are connected to the $e$ is
$\mathrm{coef}\big(\prod^{d_{\max}}_{i=1}(1+xz^i)^{\rho_i
n},x^{|e|} z^L\big)\bigg/{n \choose |e|}$.

Consequently, (\ref{Theorem2}) holds.
\end{proof}

In particular, when the degree of input node for Fountain codes is
uniformity randomly distribution,the parity matrix has constant row
weight $r$. The next theorem gives the bit erasure probability of
constant row weight ensemble.

\textbf{Theorem 3. } {\em The probability of averaged bit erasure for
Fountain ensembles $\mathcal {G}\{k,r,\rho (d)\}$ when transmitting
over a BEC with erasure probability $\varepsilon$ is}

\vspace{-0.5em}
\begin{equation}\label{Theorem2.2}
\begin{split}
&E_{\mathcal
{G}(k,r,\rho (d))}[P_b(G,\varepsilon)]\\
&=\sum_{|e|}\dbinom{n}{|e|}\varepsilon^{|e|}(1-\varepsilon)^{n-|e|}\\
&\times\sum_{L=1}^{E}\mathrm {coef}\big(\prod^{d_{\max}}_{i=1}(1+xz^i)^{\rho_i
n},x^{|e|} z^L\big)\bigg/{n \choose |e|}\\
&\times\sum^k_{V=1}\frac{V}{k}\times P(e,L,V).\\
\end{split}
\end{equation}
\normalsize \vspace{-1em}

\begin{proof} For Fountain ensembles $\mathcal {G}\{k,r,\rho (d)\}$,
similarly, if all edges incident upon a input node belong to the
edges connected to $e$, the stopping set of this input node is lost,
hence, this input node cannot be recovered.

Assuming the set of $V$ nodes connected with $L$ edges is the maximal uncorrectable set. Hence, the number of sets with $V$ input nodes and $L$ $(0\leq L\leq
E)$ edges incident upon them is

 \vspace{-1em}
\begin{flalign}
M_1(k,L,V)=&\sum_{l\leq L}\mathrm{coef}\big((1+yz^r)^k,y^V z^l
\big)(l)!.
\end{flalign}
\normalsize \vspace{-1em}

Like the proof in Theorem 2, we have
\begin{flalign}
P(e,L,V)=\frac{M(k,L,V)}{T(k,n)}
\end{flalign}
where
\begin{flalign}\nonumber
M(k,L,V)=&\Bigg(\sum_{l\leq L}\mathrm{coef}\bigg((1+yz^r)^{k},y^V z^l \bigg)(l)!\Bigg)\\
&\cdot N(k-V, E-L,0).
\end{flalign}

Hence, (11) holds.
\end{proof}

\section{Integrated Performance Analysis of Fountain Codes for BEC}
The performance of Fountain codes for BEC depends on two aspects. One is the essential codes structure, on which stopping set analysis operates. The other is the effect from the channel characteristic, which can be analyzed through the proposed uncorrectable set.

From the Theorem 6 in \cite{Shokrollahi-TIT-06-06}, the probability that $\mathcal
{G}$ has a maximal stopping set of size $s$ is at most

\begin{equation}\label{e:stopping}
S(k,\mathcal {E},s)=\dbinom{k}{s}\sum_{z=0}^n A_s(z,0)\Biggl( 1- \sum_d \rho_d \frac{\binom{n-z}{d}}{\binom{n}{d}} \Biggr)^{k-s}
\end{equation}
where $A_s(z,0)$ denotes the probability that a given subset $\Theta$ of size $s$ of the message nodes is a stopping set, given that $\Theta$ is a stopping set with $z$ check nodes of degree zero and 0 check nodes of degree one.

Eq.(\ref{e:stopping}) represents the decoding error probability due to the structure. From the above analysis, the whole decoding error set includes: (1) the uncorrectable set due to erasure, and (2) the received symbols which form a stopping set. Consequently, the final decoding error probability is

\begin{flalign}\label{e:total}\nonumber
&E_{\mathcal{G}(k,r,\rho (d))}[P_b(G,\varepsilon)]+ E_{\mathcal
{G}(k,r,\rho (d))}[S(k-V,|\mathcal {E}|-L,s)]\\\nonumber
=&\sum_{|e|}\dbinom{n}{|e|}\varepsilon^{|e|}(1-\varepsilon)^{n-|e|}\\\nonumber
&\times\sum_{L=1}^{E}\mathrm {coef}\big(\prod^{d_{\max}}_{i=1}(1+xz^i)^{\rho_i
n},x^{|e|} z^L\big)\bigg/{n \choose |e|}\\\nonumber
&\times\sum^k_{V=1}\frac{V}{k}\times P(e,L,V)\\
+&\sum_s s\times\dbinom{k}{s}\sum_{z=0}^n A_s(z,0)\Biggl( 1- \sum_d \rho_d \frac{\binom{n-z}{d}}{\binom{n}{d}} \Biggr)
\end{flalign}
where $E_{\mathcal {G}(k,r,\rho (d))}[S(k-V,|\mathcal {E}|-L,s)]$ denotes the expectation of that the $|\mathcal {E}|-L$ received nodes and $k-V$ information nodes have a maximal stopping set of size $s$.

\section{Conclusions}

This paper proposes the concept of uncorrectable sets for Fountain codes as the conventional stopping set cannot completely model the performance of the Fountain codes, especially in BECs. The advantage of the proposed mechanism is that it allows the transmission system to analyze the performance of codes when output nodes are erased. The probability of averaged bit erasure over BEC is analyzed. It can help us design efficient codes according to channel states. In the future research, we will design an algorithm with low complexity to rapidly estimate the decoding error probability.

\bibliographystyle{IEEEtran}
\bibliography{jwreference}
\end{document}